 \documentclass[final,5p,times,twocolumn]{elsarticle}

\usepackage{graphicx}
\usepackage[dvips]{color}

\usepackage{amssymb}
\usepackage{amsmath}

\usepackage{lineno}

\journal{Nuclear Instruments \& Methods in Physics Research}

\begin{document}

\begin{frontmatter}



\title{The T2K Side Muon Range Detector (SMRD)}



\def\A{\kern+.6ex\lower.42ex\hbox{$\scriptstyle \iota$}\kern-1.20ex a}
\def\E{\kern+.5ex\lower.42ex\hbox{$\scriptstyle \iota$}\kern-1.10ex e}

\author[kobe]{S.~Aoki}
\author[ox]{G.~Barr}
\author[ifj]{M.~Batkiewicz}
\author[ifj]{J.~B\l{}ocki}
\author[lsu]{J.D.~Brinson}

\author[lsu]{W.~Coleman}
\author[ifj]{A.~D\A browska}
\author[upit]{I.~Danko}
\author[wut]{M.~Dziewiecki}
\author[lsu]{B.~Ellison}
\author[inr]{L.~Golyshkin}
\author[lsu]{R.~Gould}
\author[kobe]{T.~Hara}
\author[lsu]{J.~Haremza}
\author[lsu]{B.~Hartfiel}
\author[sil]{J.~Holeczek}
\author[inr]{A.~Izmaylov}

\author[inr]{M.~Khabibullin}
\author[inr]{A.~Khotjantsev}
\author[warsaw]{D.~Kie\l{}czewska}
\author[sin]{A.~Kilinski}
\author[sil]{J.~Kisiel}
\author[inr]{Y.~Kudenko}
\author[lsu]{N.~Kulkarni}
\author[wut]{R.~Kurjata}
\author[lsu]{T.~Kutter\corref{lsu1}}

\author[sin]{J.~\L{}agoda}
\author[lsu]{J.~Liu}
\author[wut]{J.~Marzec}
\author[lsu]{W.~Metcalf}
\author[RAL]{C.~Metelko}
\author[sin]{P.~Mijakowski}
\author[inr]{O.~Mineev}

\author[upit]{D.~Naples}
\author[lsu]{M.~Nauman}
\author[RAL]{T.C.~Nicholls}
\author[upit]{D.~Northacker}
\author[lsu]{J.~Nowak}
\author[IC]{M.~Noy}
\author[upit]{V.~Paolone}
\author[RAL]{G.F.~Pearce}
\author[lsu]{O.~Perevozchikov}
\author[warsaw]{M.~Posiada\l{}a}
\author[sin]{P.~Przew\l{}ocki}
\author[RAL]{W.~Qian}

\author[IC]{M.~Raymond}
\author[lsu]{J.~Reid}
\author[sin]{E.~Rondio}
\author[inr]{E.~Shabalin}
\author[RAL]{M.~Siyad}
\author[lsu]{D.~Smith}
\author[itp]{J.~Sobczyk}
\author[ifj]{M.~Stodulski}
\author[sin]{R.~Sulej}
\author[ifj]{J.~\'Swierblewski}
\author[kobe]{A.T.~Suzuki}
\author[sil]{T.~Szeg\l{}owski}
\author[sin]{M.~Szeptycka}
\author[RAL]{M.~Thorpe}
\author[ifj]{T.~W\A cha\l{}a}
\author[csu]{D.~Warner}
\author[RAL]{A.~Weber}
\author[kobe]{T.~Yano}
\author[inr]{N.~Yershov}
\author[ifj]{A.~Zalewska}
\author[wut]{K.~Zaremba}
\author[wut]{M.~Ziembicki}


\address[IC]{Imperial College London, Physics Department, London, United Kingdom}
\address[inr]{Institute for Nuclear Research, Moscow, Russia}
\address[csu]{Colorado State University, Department of Physics, Fort Collins, Colorado, USA}
\address[ifj]{H. Niewodnicza\'nski Institute of Nuclear Physics PAN, Krak\'ow, Poland}
\address[kobe]{Kobe University, Department of Physics, Japan}
\address[lsu]{Louisiana State University, Department of Physics \& Astronomy, Baton Rouge, Louisiana , USA}
\address[sin]{National Centre for Nuclear Research, Warsaw, Poland}
\address[ox]{University of Oxford, Department of Physics, Oxford, United Kingdom}
\address[upit]{University of Pittsburgh, Department of Physics and Astronomy, Pittsburgh, Pennsylvania, USA}
\address[sil]{University of Silesia, Institute of Physics,  Katowice, Poland}
\address[RAL]{STFC Rutherford Appleton Laboratory, Harwell Oxford,  United Kingdom}
\address[warsaw]{University of Warsaw, Institute of Experimental Physics, Warsaw, Poland}
\address[wut]{Warsaw University of Technology, Institute of Radioelectronics, Warsaw, Poland}
\address[itp]{Wroc\l{}aw University, Institute of Theoretical Physics, Wroc\l{}aw, Poland}


\begin{abstract}

The T2K experiment is a long baseline neutrino oscillation experiment aiming 
to observe the appearance of $\nu_e$ in a $\nu_\mu$ beam. The $\nu_\mu$ 
beam is produced at the Japan Proton Accelerator Research Complex (J-PARC),
observed with the 295~km distant Super-Kamiokande Detector and monitored by
a suite of near detectors at 280m from the proton target. The near detectors 
include a magnetized off-axis detector (ND280) which measures the un-oscillated 
neutrino flux and neutrino cross sections. The present paper describes the
outermost component of ND280 which is a side muon range detector (SMRD) 
composed of scintillation counters with embedded wavelength shifting fibers 
and Multi-Pixel Photon Counter read-out. The components, performance 
and response of the SMRD are presented.
\end{abstract}


\end{frontmatter}


\section{Introduction}
\label{intro}

The Tokai to Kamioka (T2K) experiment is a second generation long-baseline accelerator neutrino oscillation experiment\cite{t2knim,ito,mac}
which started taking data in January 2010.
The principal goal of the project is to measure neutrino oscillation parameters by searching for the appearance of electron neutrinos ($\nu_{e}$)
in a beam of muon neutrinos ($\nu_{\mu}$) and by performing precision measurements of the muon neutrino disappearance.
The project utilizes protons from the new  30~GeV Main Ring (MR) proton synchrotron at J-PARC in Tokai\cite{mac} to produce a $\nu_{\mu}$ beam on a 
stationary graphite target.
A 295~km distant detector at Kamioka is placed 2.5$^\circ$ off-axis to measure beam neutrinos. The upgraded water Cherenkov detector, 
Super-Kamiokande\cite{t2knim,sknim}, serves as far detector for T2K.
The use of an off-axis beam results in a neutrino energy spectrum that is narrowly peaked at about 600 MeV corresponding to the oscillation maximum 
for a baseline of 295 km. The off-axis configuration also leads to a significant reduction in backgrounds originating from a beam intrinsic 
electron neutrinos. The estimated fraction of electron neutrinos at the peak energy is $\nu_e/\nu_{\mu}\sim0.4\%$. 
A near detector complex at J-PARC is used to monitor the neutrino beam, measure neutrino event rates and to help minimize systematic uncertainties 
in the measurements of neutrino oscillation parameters. A detector facility is located 280~m downstream from the target
was built at J-PARC in Tokai for this purpose.
Neutrino oscillation parameters can be derived from flux and spectral measurements of ${\nu_{\mu}}$ and ${\nu_e }$ performed with the Super-Kamiokande 
and with the near detectors. The near detectors provide a normalization of the unoscillated flux and a measurement of backgrounds.
The near detector complex 
houses two detectors\cite{le,kud}: 
an on-axis interactive neutrino grid detector (INGRID)  
and an off-axis near detector, ND280 \cite{t2knim,nd280}.
Protons are accelerated up to 30~GeV and proton spills are successively 
extracted every 2.6 seconds (initially 3.5~s) from the J-PARC main ring synchrotron 
into the neutrino beamline. The neutrino beamline leads to a graphite 
target\cite{mat} onto which protons impinge.  
The target is embedded in the first of three magnetic horns which are excited  by a pulsed current that is synchronized with the beam.
During (anti-)neutrino runs the horns serve to focus (negatively)positively 
charged pions into a 96~m long decay volume, where they decay into muons and muon neutrinos.\\
The neutrino beam direction needs to be tuned and monitored
precisely since for an off-axis configuration the peak neutrino energy 
changes by 2\% per millirad change in beam direction. 
The neutrino beam flux and direction is measured  by the on-axis INGRID  detector which is composed of an array of iron-scintillator sandwich detectors.
The off-axis detector ND280 is positioned at a 2.5$^\circ$ off-axis angle as is the far detector and it's primary purpose is to meausre the unoscillated
neutrino energy spectrum, the $\nu_e$ contamination in the beam and neutrino interaction cross sections.
The dominant neutrino interaction process around the T2K neutrino peak energy of about 600~MeV is charged-current quasi-elastic (CC-QE) scattering
$\nu_{\mu}n\rightarrow\mu^{-}p$.
The CC-QE cross-section measured by ND280  will be used as a reference cross-section for  measurements  in the Super-Kamiokande detector.
In order to achieve the projected sensitivities \cite{ito} of the experiment, the momentum resolution of muons from CC-QE  interactions 
should be better than 10\%\cite{ito}.
The expected precision  for determining  the energy scale is better than 2\% and the 
backgrounds for a $\nu_e$  appearance should be measured with  uncertanties smaller than 10\% .

\section{The Off-axis Detector and the SMRD}

\begin{figure}[htp]
\begin{center}
\includegraphics[width=0.45\textwidth]{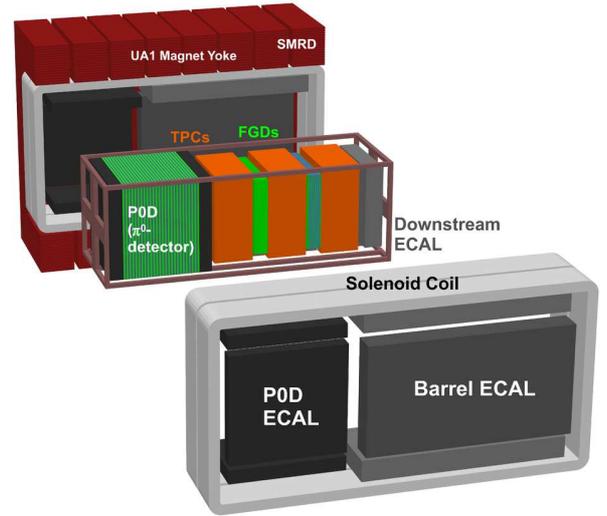}
\caption[]{A view of the ND280 off-axis detector.}
\label{fig:nd280_1}
\end{center}
\end{figure}

The off-axis near detector ND280 consists of finely segmented  detectors which act as neutrino targets and tracking detectors and are 
surrounded by an electromagnetic calorimeter (ECal) and a  magnet which houses a side muon range detector (SMRD).
The inner part of the detector is divided into multiple regions: a $\pi^0$ detector (P0D), a tracking detector which in turn consits of
two fine grained detectors (FGD) made from scintillators and water modules and three gaseous time projection chambers (TPC).
The layout of the detector is shown in Fig. \ref{fig:nd280_1}. 

The P0D detector  consists of scintillator planes interleaved with lead, brass and water layers.
It is designed to measure backgrounds for the electron neutrino appearance search in the far detector.
In particular the background  from the neutral current reaction with only a single  $\pi^0$ in the final state
and the electron neutrino contamination in the beam.
Segments with  water are implemented for neutrino  water cross-section measurements.\\
Downstream of the P0D is the tracking detector which is optimized to measure momenta of charged particles, 
particularly muons and electrons and recoil protons produced in charged current quasi-elastic interactions.
The tracking detector consists of two fine grained detectors which are interspersed with three time  projection chambers.
The FGDs provide the target mass for neutrino interactions in the tracker.  Directions and ranges  of short tracks such as recoil protons produced by 
CC-QE interactions in the FGD are measured. The first FGD module is made of 
layers which in turn are composed of scintillator bars with
perpendicular orientation from one layer to the next. 
The second module consists of equivalent scintillator bars which are 
interspersed with passive water layers to
allow for cross section measurement on water. 
The primary  purpose of the TPC is to measure muon and electron momenta for neutrino energy reconstruction in CC-QE interactions  and 
ionization energy losses, $\frac{dE}{dx}$, for particle identification.\\ 
Both P0D and tracker are surrounded by an electromagnetic calorimeter to detect showering ($e^{-}$, $\gamma$) particles which escape from the tracking 
detector and P0D and allow for better event reconstruction and particle identification.
The inner part of the off-axis detector is placed inside the UA1 dipole magnet, which is operated to produce a uniform horizontal magnetic field of 0.2 T. 
The  magnet yoke  is divided into two halves, each made of eight  C-shaped flux return yokes. Each of these individual yokes consists of
sixteen  iron plates, 48 mm thick, and spaced with 17 mm air gaps.\\ 
The side muon range detector consists of a multiple layers of plastic scintillation counters placed in the air gaps in between the iron plates which make up the yokes 
of the UA1 magnet. The main purpose of the SMRD is  to measure muon momenta for muons which
are created in neutrino interactions and which escape the inner detectors at large angles with respect to the neutrino beam.
It also serves to identify backgrounds from beam neutrino interactions in the magnet yoke and surrounding walls and to provide a cosmic trigger signal 
for calibration purposes of the ND280 detector. A detailed description of the ND280 near detector  can be found in \cite{t2knim}.\\
One of the  main goals of the ND280 detector is to measure the neutrino energy spectrum. For charged current quasi elastic CC-QE processes,  
neutrino energy is closely related to the muon momentum and its angle of emission. The active target mass  is concentrated in the FGD and P0D detectors. 
Muons emitted at large angles with respect to the neutrino beam direction often leave short or no tracks in the TPC.
Such muons will intersect the ECal as well as the SMRD and the iron yokes surrounding the entire detector. 
Their momenta can be inferred from the range in iron and the SMRD measurements.
The directions of muons can be determined based on information from the inner detectors and the position information in  the plastic scintillation 
counters of the SMRD.\\
Estimates based on MC studies indicate that a significant fraction
of all muons which originate from CC-QE reactions intersect the SMRD.
The vast majority of large angle muons which does not escape in the forward direction, e.g. for $cos ~\theta > 0.8$, have momenta of less than 600~MeV/c.
Muons with energies less than   600~MeV will range out within less than 350~mm of iron. The iron plates of magnet yokes are 48~mm thick 
and hence it is sufficient  to instrument 6 to 7 radial layers in order to range out the majority of muons that are not escaping in the forward direction.

The principal requirement for the SMRD detector is to detect particles with very high efficiency. Hence the active detector medium  has to enclose the 
inner detectors nearly hermetically and provide uniform efficiency across the entire sensitive area.  An additional requirement is the long-term stability 
of all detector components to ensure good performance and longevity over the lifetime of the experiment.
The SMRD will also be used  to identify backgrounds entering the detector from outside as well as from interactions of beam neutrinos in the iron of the 
magnet yoke.
The SMRD  provides the trigger signal in response to through-going  cosmic ray muons. These muons are an invaluable calibration source for the 
inner detectors  as they provide a sample of muon tracks that are, apart from their direction, very similar to the muons created in neutrino beam 
interactions.

\section{SMRD Hardware}

The side muon range detector is embedded in the former UA1 and NOMAD magnet which is now located at J-PARC in Tokai, Japan. The magnet is housed in a 
17~m deep detector pit without overburden, 280m downstream from the carbon target and at an off-axis angle of 2.5$^\circ$. 
The following sections describe the geometry, design and components of the SMRD.

\subsection{The ND280 Off-Axis Magnet}

\begin{figure}[htp]
  \begin{center}
    \includegraphics[width=0.25\textwidth]{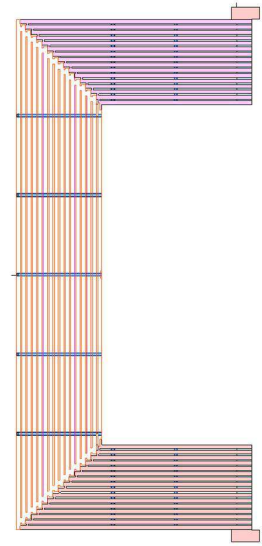}
  \end{center}
  \caption{Engineering drawing view of a single yoke. A series of bolts hold the iron plates together and sub-divide each yoke into 4 horizontal, 4 vertical and 2 corner sections.}
  \label{fig:yoke}
\end{figure}

The former UA1 magnet consists of 16 flux return yokes which are grouped in pairs (labeled 1 through 8 from upstream to downstream with respect to the neutrino beam direction) to form a series of 8 consecutive rings which surround the magnetic field and the current coils on four sides. All 16 yokes have essentially the same dimensions with the four yokes forming the first and the last ring exhibiting slightly different geometry at the interface where two yokes meet to form a ring. The nominal outer dimensions of each yoke are 6150mm $\times$ 2815mm $\times$ 876mm (height $\times$ width $\times$ depth) and the enclosed space measures 4040mm $\times$ 3600mm $\times$ 7568mm (height $\times$ width $\times$ depth).
 The relative ring spacing amounts to 110~mm for the outermost rings and 80~mm for all other rings. 
The original UA1 inner volume width was 80mm smaller but the yokes were slightly modified for the NOMAD experiment.
Each yoke is composed of sixteen 48mm thick iron plates which are separated by 17mm thick spacers and held together by long bolts. Figure \ref{fig:yoke} shows the basic structure of a single yoke and the slits in between iron plates.
Slits which are located in between the same set of bolts form a group labeled as tower.
Except for corner towers, all horizontally oriented slits have a nominal gap size of 17mm $\times$ 700mm $\times$ 876mm (height $\times$ width $\times$ depth),
and all vertically oriented slits have a nominal gap size of 910mm $\times$ 17mm $\times$ 876mm (height $\times$ width $\times$ depth). Due to imperfections and deformations of the magnet as well as welding seams of the spacers the actual gap sizes may differ slightly.

\subsection{The SMRD}

Each yoke consists of 16 iron plates which are spaced at a distance of 17mm which leads to a set of 15 slits in radial direction. 
Active detector components, populate the three innermost gaps in all yokes so as to be able to detect particles escaping the inner detectors .
Particles which escape the inner detectors on the sides have slightly higher mean energy compared to particles which are exiting through the
top or bottom where the magnet coil material causes additional energy loss.
Therefore, additional active layers have been installed in the sides of the most downstream yokes; a total of 4 vertical layers in yoke 6 and 
6 layers in yokes 7 and 8. The location of the larger number of layers of active detector material was chosen to maximize the the observation 
of muons originating from beam related neutrino interactions inside ND280.

The SMRD consists of 192 horizontally oriented and 248 vertically oriented detector modules in total. The external dimensions
of the detector modules are dictated by the dimensions of the slits in the yokes and measure to
9mm $\times$ 686mm $\times$ 955mm (height $\times$ width $\times$ depth) for horizontal and 
9mm $\times$ 892mm $\times$ 955mm (height $\times$ width $\times$ depth) for vertical modules.
The module height is pointing radially outward and the module depth is parallel to the long axis of the magnet.

The counter sizes have been optimized to maximize the active area in each magnet slit. Due to the differently sized spaces for horizontal and vertical slits, horizontal SMRD modules are composed of 4 scintillation counters with dimensions 7mm $\times$ 167mm $\times$ 875mm (height $\times$ width $\times$ length) and vertical SMRD modules consist of 5 scintillation counters with dimensions 7mm $\times$ 175mm $\times$ 875mm (height $\times$ width $\times$ length). Hence, the SMRD consists of 768 horizontal and 1240 vertical scintillation counters in total.

\begin{figure}[htp]
\begin{center}
\includegraphics[width=0.45\textwidth]{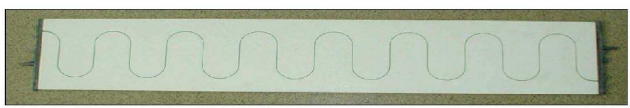}
\caption{SMRD scintillator slabs with a serpentine-routed Y11 WLS fiber.}
\label{fig:smrd_counter}
\end{center}
\end{figure}

Polystyrene-based scintillator slabs, each with an embedded wave-length 
shifting (WLS) fiber, serve as active elements for the SMRD detector. 
The WLS fiber is read out on both ends to increase light yield, improve
uniformity and position accuracy, and provide redundancy (Fig.~\ref{fig:smrd_counter}). 
A key feature of the individual  SMRD counters is the usage of a serpentine-shaped fiber. 
The design results in near uniform response across the surface of the scintillation counter \cite{mineev} and 
minimize the number of photosensors and electronics channels compared to more conventional designs with multiple straight fibers.
   Extruded polystyrene scintillators (1.5\% PTP, 0.01\% POPOP) have been produced by Uniplast, a company in Vladimir, Russia. 
The outer surfaces of each slab are etched by a chemical agent, thus resulting in formation of a white diffuse layer. 
The deposit thickness is 30-100~$\mu$m depending on etching time. The diffuse layer acts as a reflector and has been demonstrated to 
increase the light yield by 40 to 50\% relative to a scintillation counter without such a diffusively reflective layer.
The advantage of the approach is an almost ideal contact of the reflector with the scintillator. 
Details about the extrusion technique and the method of etching a scintillator with a chemical agent can be found in \cite{extrusion}. 
Extruded scintillators of this type have shown good light yield stability over at least two years \cite{scint_stability}.
   After the etching procedure raw scintillator slabs were cut with a milling machine and the serpentine-shaped  groove was milled successively.
For the SMRD counters the serpentine geometry of a groove consists of 15 half-circles, each with a diameter of 58~mm and straight sections connecting
the semi-circles. Due to the different widths of horizontal and vertical counters grooves of two different lengths are required: 
2.10~m for horizontal counters and 2.22~m for vertical ones. The groove depth is 2.8~mm and deepens to 3.55~mm towards both ends of the counters 
so as to allow the fiber to exit the scintillator at mid-height. Grooves were milled in multiple passes with 
a 1.1~mm diameter mill to ensure good optical surface properties of the grooves. The accuracy of the milling operation was 100~$\mu$m.  
 A 1~mm diameter Y11 (150) Kuraray WLS fibers of flexible S-type and with double-cladding \cite{kuraray} was used for the SMRD counters.
Fibers are bent into a serpentine-shape and glued into grooves with BC600 Bicron glue. Details of the gluing technique are described in \cite{bicron}.\\
The long term stability of bent WLS fibers was tested with two photosensors MRS APDs \cite{mrs_apd} and a blue LED as a light source. 
Several 3~m long fiber samples were wound into 7 turns to reproduce similar bending stresses.
A bending diameter of 60~mm led to an average initial drop in light transmission quality of about 5\%, 
which is in a good agreement with Kuraray data \cite{kuraray}.  
The light attenuation was measured as the ratio of the light signal at the far fiber end to the signal at the close end. 
The near end signal served as a reference of LED intensity. A straight fiber of the same length was used to calibrate the photosensors. 
No degradation in light attenuation other than the initial decrease has been observed over a period of more than a year.

\begin{figure}[htp]
  \begin{center}
    \includegraphics[width=0.4\textwidth]{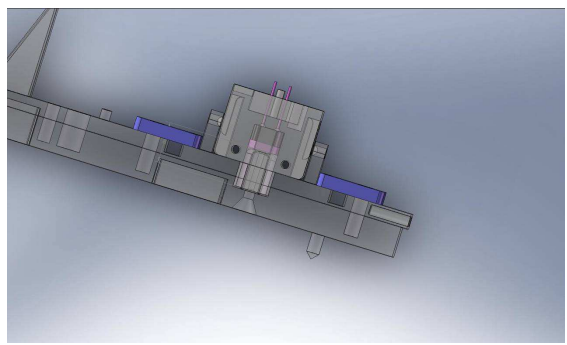}
  \end{center}
  \caption{Partial view of a SMRD scintillator counter endcap with the optical connector housing a MPPC and a mini PCB.}

  \label{fig:endcap}
\end{figure}

%
In the SMRD scintillation counters
the WLS fiber exits both sides of the scintillator through a ferrule which is part of a custom made endcap.
Endcaps are glued and screwed to the end faces of each scintillator counter, and are injection molded and made out of black vectra, a brand of liquid crystal polymer and 
which was tested to be light tight at the few photon level.
Each endcap features a snap-on mechanism to engage a connector with an inserted Hamamatsu multi pixel photon counter (MPPC) \cite{mppc}, to couple to the endcap and the WLS fiber ferrule, respectively. The MPPCs are backed by a foam spring to ensure a reliable coupling between the photosensor front face
and the polished WLS fiber ends.
Figure \ref{fig:endcap} shows an engineering drawing of the connector containing a photosensor and clipped to the endcap.
\\
All 4016 MPPCs of the SMRD are connected to miniature printed circuit boards (PCB) which are free to slide along rails in the backside of the optical connector.
The free movement of the PCB along the leads of the photosensors ensures good optical connections between WLS fiber and MPPC and 
averts damage to the MPPC front face due to overpressure. 
The miniature PCBs couple the MPPC signals into Hirose mini-coax cables which lead the signals to the Trip-t frontend boards 
(TFBs) \cite{tript,tfb}.
The front end electronics is mounted on the vertical sections of the magnet yokes.
The mini coax-cables are routed between the magnet yokes and measure 3.5m and 3.0m for horizontal modules and are 2.2m long for all vertical modules.

\subsection{Photosensors}

Multi Pixel Photon Counters (MPPC) \cite{mppc} developed for T2K by Hamamatsu were chosen as the common photosensor for all ND280 scintillator based detectors.
Key features of these devices are their insensitivity to magnetic fields and their compact size which make them well suited for applications near or in
the magnetic field and within the limited space available inside the UA1 magnet. The total number of MPPCs used for the SMRD amounts to 4016.

The custom made version of the MPPCs for T2K consist of an array of 667 independent  $50\times50~ \mu m^2$  avalanche photodiodes (pixels) 
operating in Geiger mode. The MPPC sensitive area of $1.3\times1.3~mm^2$ is well suited to accept light from a 1~mm diameter Y11 fiber. 
The MPPC signal is a sum of pixel avalanches and this multi-pixel sensor operates as an analog photodetector with a dynamic range that is limited by 
the finite number of pixels. Typical light signals in SMRD counters are below 50~photoelectrons and therefore dynamic range issue are not a concern.
Each pixel can be represented as a microcapacitor which quickly discharges during Geiger breakdown initiated by a photoelectron until the voltage 
difference across it has decreased below the breakdown voltage. The overvoltage, which is defined as the difference between the supplied bias voltage 
and the breakdown voltage, is the main parameter that affects the performance of MPPCs and the stability of its operation.  
MPPCs have an excellent single photoelectron resolution up to mean charges corresponding to about 30 photoelectrons
that allow to perform an accurate calibration of each counter. \\
At a temperature of $T=25^{\circ}C$ and an overvoltage of 1.6~V MPPCs are characterized by a typical gain of $7.5\times10^5$, a photo detection efficiency 
of about 25\% for green light as emitted by a Y11 fiber. The average dark rate
 amounts to 700~kHz with maximum values approaching up to 1~MHz,
the estimated combined crosstalk and afterpulse probability is 20-25\% and the 
recovery time of a single pixel is 30~ps. 
The MPPCs of the SMRD were operated in the T2K neutrino beam starting in 2009 and after more than 1.5 years of operation
only one sensor (0.025\%) is suspected to have failed.
All MPPCs were tested extensively as function of bias voltage and temperature and in particular the gain and dark rate had to satisfy stringent criteria
in order for a MPPC to be included in the SMRD.

\subsection{Module Assembly and Installation}
\label{assembly}

At multiple stages of the detector production and assembly the performance of the scintillation counters with embedded WLS fiber were tested in response to cosmic rays. 
First, the scintillators were tested immediately after the extrusion 
process by measuring the light yield with a photomultiplier tube and in response to throughgoing muons. 
Secondly, after the endcaps were attached to the scintillation counters and the WLS fiber had been glued into the grooves with BC600 optical glue the counters were retested using MPPCs and double ended readout in responsee to central penetrating muons.
Out of 2008 counters 20 were found to have a large asymmetry (more than 50\%) in light yield between the two ends. The asymmetry was attributed to a
damages of the fiber cladding encurred during the gluing process. These 20 counters were repaired by gluing a new fiber into a refurbished groove
after milling out the previously glued fiber. All counters which passed the 
quality test, were wrapped by one layer of 0.1~mm thick Tyvek paper which leads to a
further increase in light yield of 15\%.
\begin{figure}[htb]
\centering\includegraphics[width=0.5\textwidth]{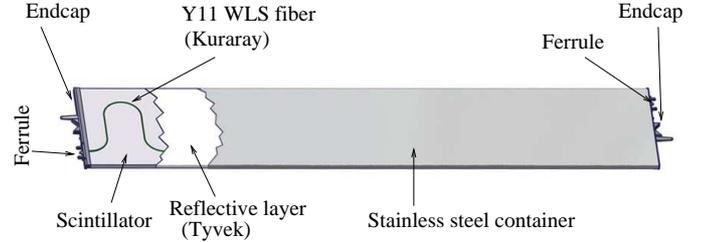}
\caption{SMRD counter sliced view.}
\label{fig:counter_sliced}
\end{figure}

Successively the counters were wrapped in a lightproof stainless steel container 
(Fig.~\ref{fig:counter_sliced}). The container was attached to scintillator and endcaps 
with DP-490 black epoxy glue and a double sided 20$\times$0.15~mm Tesa adhesive tape. 
Additionally all joints between the container surface and the endcap were covered by a black Tesa adhesive tape of 25$\times$0.065~mm. Each fiber end inside the endcap ferrule 
was cut by a cylindrical mill and polished to provide good optical contact with the MPPC. After assembly, the  dark noise of each SMRD counter was measured with MPPCs and 
an oscilloscope to ensure the absence of light leaks.
In total, 2130 counters (800 -- 167~mm wide and 1330 -- 175~mm wide counters) were assembled and 
tested to be of good quality.
After shipment to Japan 230 counters were found to have developed a sub-millimeter sized
air gap between the end of the fiber and the face of the ferule, resulting in a small 
loss in light yield.
Hence the endcaps of all counters were additionally fastened by 2 stainless steel 
screws each in order to minimize the risk of future counter degradation.
After refurbishment of the problematic counters all counters were re-tested and 
demonstrated to show excellent performance.

\begin{figure}[htb]
\centering\includegraphics[width=0.5\textwidth]{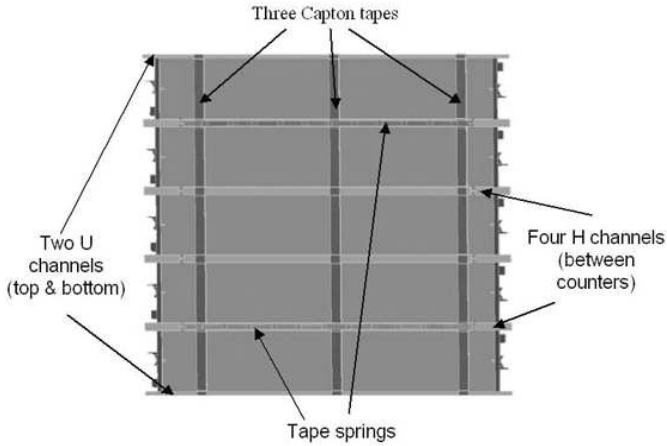}
\caption[]{Scheme of the vertical  SMRD module.}
\label{fig:mod-1}
\end{figure}

\begin{figure}[htb]
\centering\includegraphics[width=0.5\textwidth]{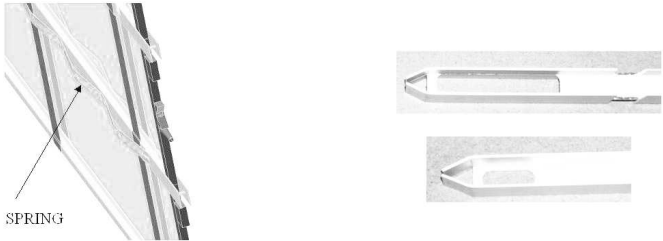}
\caption[]{Springs mounted and fixed on H–profiles (left), tapered end of C and H-profiles (right).}
\label{fig:mod-2}
\end{figure}

\begin{figure}[htb]
\centering\includegraphics[width=0.4\textwidth]{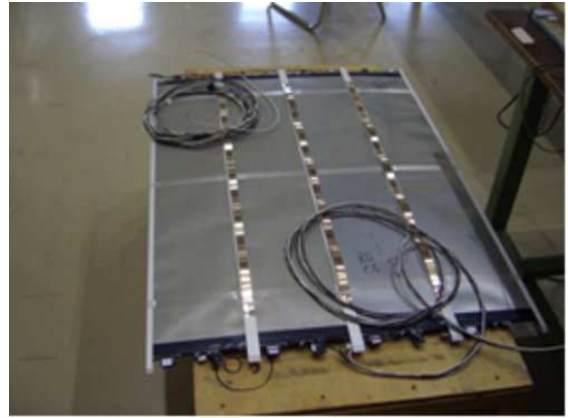}
\caption[]{Completed SMRD horizontal module equipped with photo-sensors and combined power and signal cables.}
\label{fig:mod-3}
\end{figure}

Single counters are assembled into bigger units named modules to facilitate installation and to stabilize the position 
of the counters in the magnet slits. In order to match the different dimensions of vertical and horizontal magnet slits
two types of SMRD modules were built.
Modules intended for vertical slits consist of five counters (each 175mm wide)
while the horizontal ones consist of four counters (each 167 mm wide).  
Extruded Aluminum H-profiles are used to inter-connect counters into modules. 
The boundary edges of the first and last counter in each module were protected with
aluminium U-channels as shown in fig. \ref{fig:mod-1}.
The counters and the extrusions are tightly wrapped with capton tape in three locations. 
In order to stabilize a module inside a magnet slit
tape springs made of phosphor–bronze are mounted on both sides of the modules as indicated in Fig.~\ref{fig:mod-2}. 
Two springs are mounted on each side of the vertical modules while three per side are attached to the H-profiles
of horizontal modules.
Any lateral and longitudinal forces from the springs act on the H-profiles and not on the counters or endcaps
whose interconnection is considered to be the most sensitive part of a detector module. 
The springs are 900 mm long, 15 mm wide, 0.4 mm thick. Each tape spring has 5 waves. 
The shapes of the springs are symmetric to ensure proper module positioning during installation
in the magnet. 
Both ends of each aluminium extrusion are tapered (right panel of Fig.~\ref{fig:mod-2}) to facilitate the installation process.
The final assembly and installation of the 248 vertical and 192 horizontal SMRD modules was performed at J-PARC between February and June 2009. 
A photo of a SMRD horizontal module is shown in Fig.~\ref{fig:mod-3}.

SMRD modules were inserted into the magnet slits from the front and back face 
of the magnet and slid to their final positions.
Each module was tested for possible damages during installation by reading out 
photosensor signals immediately after positioning it in its final location.
After the completion of installation and checks 
protective covers were installed on the front faces of the magnet to prevent 
any damages to SMRD modules.

\subsection{Counter Quality Assurance and Pre-installation Tests}
\label{section:qa}

\begin{figure}[htb]
\begin{center}
\includegraphics[width=0.45\textwidth]{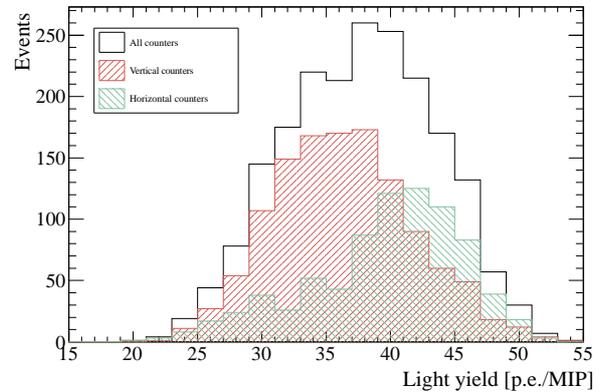}
\caption{ The light yield distribution for 2024 SMRD counters (including 16 spares).  }
\label{fig:ly_qa}
\end{center}
\end{figure}

Quality assurance (QA) of all SMRD counters was carried out at multiple stages throughout the production process.
The light yield of raw scintillator slabs was tested in response to penetrating cosmic muons and with a photomultiplier attached for readout.
After machining the S-shaped groove and gluing of a WLS fiber counters were tested with cosmic muons passing through an area of 7$\times$7 cm$^2$ 
in the central area. In order to assure good response of the counters for an anticipated experiment lifetime of 10 years
the light yield requirement for the sum of both ends measured at 20$^\circ$C is more than 25 p.e. per minmum ionizing particle (MIP).
Quality assurance measurements show the light yield to range from 25 to 50 p.e./MIP (Fig.~\ref{fig:ly_qa}). 
For centrally penetrating muons differences in signal sizes at both counter ends are required to be less than 20\% so as to guarantee near uniform 
response and no fiber damage. Less than 1\% of the counters were rejected but passed the quality assurance after a new WLS fiber had been inserted (see 
section \ref{assembly}). 

\begin{figure}[htb]
\begin{center}
\includegraphics[width=0.45\textwidth]{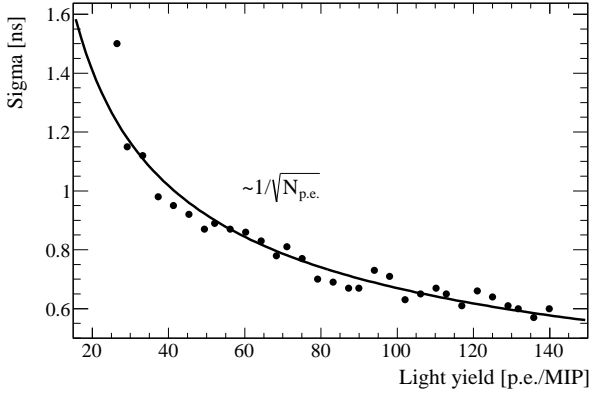}
\caption{Time resolution of a SMRD counter versus light yield (after time-walk correction). The data are recorded in response to cosmic muons 
penetrating the center of the scintillator slab. }

\label{fig:timesigma_ne}
\end{center}
\end{figure}

The performance of all SMRD counters with attached MPPC photo-sensors was measured again after shipment and assembly into modules.
A small 2 $\times$ 2 cm$^2$ muon telescope was placed at the center of each counter and used to trigger 
signal read-out on both ends  of the SMRD counters.
The mean light yield for the sum of both ends was near 40 p.e./MIP after subtraction of MPPC cross-talk and after-pulses.  
The mean detection efficiency for mimimum ionizing particles was in excess of 99.9\%. 
The temperature during the performance tests was in the range from 20.5 to 
22.5$^\circ$C and the bias voltages of both MPPCs were set to 68.7 V.  
In addition, the timing resolution of counters, defined as the uncertainty on $(T_{left}-T_{right})$/2 was measured as function 
of light yield (Fig.~\ref{fig:timesigma_ne}). 
For a light yield of 40 p.e. a time resolution of 0.9 ns was achieved. 
The position resolution of a counter along its major axis based on both time difference between signals at both counter ends and 
the charge asymmetry between the signals was estimated and found to be of 
order of 8.5 cm.

\subsection{Electronics and Power Requirements}

The electronics for the SMRD is based on the TRIP-t 
ASIC
developed at Fermilab\cite{tript}. 
It is the same electronics as used by some of the other 
sub-detectors of the T2K near detector. For an overview of the 
near detector complex and its components see \cite{t2knim}.

Signals from up to 64 MPPCs are routed to custom designed front end boards
(TFB: TRIP-t Front-end Board)
that house 4 TRIP-t ASICs using miniature coax
cables.  The signals from the photosensors are capacitively split
and routed to separate channels of the ASIC to increase the dynamic
range of the electronics.  A one photo-electron (p.e.) signal corresponds to
around 10 ADC counts in the high gain channel, while the full scale
signal in the low gain channel corresponds to 500 pe.

\label{chapter:triggerlogic}

The TRIP-t chip is integrating the charge in programable integration
windows, which are synchronized with the neutrino beam structure. There
is a programmable reset time after each integration cycle, which is at
least 100 nsec long. The chip can store the result of 23 integration
cycles in a capacitor array. Once the 23 integration cycles have been
recorded, the data is multiplexed onto two dual channel 10-bit ADCs,
which digitize the data. Signals from the high gain channel are routed to
a discriminator, which is part of the TRIP-t ASIC. The front-end board
is controlled by an FPGA, which timestamps the data with a precision of
2.5 ns. The threshold to generate a timestamp is programmable from 0 to
5 p.e.  The ADC and timestamp data is assembled by the FPGA and send to
a backend board for data concentration and buffering. The output of the
discriminators are also used to calculate trigger primitives, which are
used to initiate the readout of the detector for cosmic ray muons. One
normally requires hits from both ends of a scintillator counter within a
23 ns coincidence window and at
least 2 counters to be hit, before a trigger primitive is transmitted to
the cosmic trigger module.  Monitoring information of temperature
and MPPC bias voltages are also recorded by the TFB and asynchroniously transmitted to
the RMM. More detail of the front end part of the electronics can be
found in
\cite{tfb}.

The backend of the electronics system consists of four readout merger
modules (RMMs), a cosmic trigger module (CTMs), one slave clock
module (SCM) and a master clock module (MCM). All the boards have
a common hardware platform, which has been build around a high end Vertex II
Pro FPGA from Xilinx, which is clocked at 100 MHz. 
The board can drive 14 high speed optical links
via its RocketIO and up to 192 LVDS links.

The signals from 128 TFBs, which are mounted on the detector and up to 
3.5 m away from the photosensors, are routed to the four RMMs via cat5e
cables. Each RMM controls 32 TFBs, distributes the clock and trigger
signals and receives the data after a trigger signal was received by
the TFBs. It asynchroniously sends data via Gigabit Ethernet
link to a commercial PC that collects and processes the data. Each RMM is
equipped with 500 MByte DDR2 memory and can buffer up to 128 triggers.
The RMMs receive trigger and timing signals from the SCM.

The MCM receives signals from the accelerator that
determine when the neutrino spill happens and also from a GPS based
clock. The GPS signals are used to synchronize the electronics to
coordinated universal time (UTC). 
The MCM is also connected to the cosmic trigger module (CTM), which
receives signals from up to 192 TFBs. All 128 TFBs used to read out the
SMRD are set up to contribute to the generation of
cosmic triggers. Readout is initiated, if
at least 2 TFBs reading out data on two different sides of the detector
are generating a trigger primitive in a 200 nsec window.  
The CTM is highly configurable. The trigger primitives can be prescaled, 
to suppress some combinations of SMRD walls or towers, e.g. to allow
a cosmic muon event sample with preferentially horizontal directions. 
The prescale factors are set using an 8-bit word.
All timing and trigger signals are transmitted via a RocketIO driven optical 
link from the MCM to the SCM.\\

The power requirements for the TFBs and RMM/CTMs are listed in table~\ref{table:TFB-RMMpwr}.  
Eight TFBs are mounted on the vertical side of each magnet yoke.
All TFBs are mounted, evenly spaced, onto 6063 Aluminum 
architectural C-Channel that spans the entire vertical height of the magnet yokes.
The C-channel is electrically isolated from the magnet steel to reduce the possibility
of noise pickup. In addition to structural support for the TFB mounting, the inside of 
the C channel itself is used for power cable routing and strain relief.

Power to the TFBs is distributed by a system of power distribution panels. Two
primary power panels service one half of the magnet each. These panels are 
mounted on lower middle section of the magnet and are directly
connected to two Wiener PL508 power supplies located on a lower level.
One primary panel connects via 10 AWG wire to 8 secondary 
power distribution panels which are mounted on the bottom of each yoke and 
C-Channel mounting rail. 
Each secondary power panel services 8 TFBs and the power to 
each TFB is supplied using 22 AWG wire. The voltages for each secondary panel are fused: 5~A fuse 
for 1.7~V, 7.5~A fuse 3.1~V, 1~A fuse for 3.8~V and 2~A fuse for 5.5~V. 
In case of an electrical short on one of the TFBs 
at most 8 TFBs will lose voltage. The power to the 4 RMMs and CTM are supplied and 
fused (7.5~A fuse for both 3.1~V and 3.8~V and 1~A fuse for 5.5~V) at the primary power distribution panels. 
\begin{table}[ht]
\begin{center}
\begin{tabular}{|c|c|c|}
\hline
\multicolumn{3}{|c|}{Power Requirements} \\
\hline
\hline
Voltage(V) & \multicolumn{2}{|c|}{Current(A)} \\
 & per TFB & per RMM/CTM \\

\hline
1.7 & 0.24 & -- \\
3.1 & 0.60 & 1.0\\
3.8 & 0.05 & 6.0\\
5.5 & 0.18 & 6.0 \\
\hline
\end{tabular}
\caption{Power requirements per TFB and RMM or CTM. The SMRD uses a total of 128 TFBs, 4 RMMs and 1 CTM.}
\label{table:TFB-RMMpwr}
\end{center}
\end{table}

\subsection{Data Acquisition System}

The SMRD is configured and read out by a global T2K ND280 data acquisition (DAQ).
It is based on the MIDAS DAQ framework \cite{midas}
operating on commercially available computing hardware running the Scientific Linux operating system.
MIDAS is interfaced to the experimental hardware
through custom C/C++ front-end client applications.

The SMRD electronics is read out and controlled by the DAQ front end processor nodes (FPN)
which are interfaced to the back-end electronics modules (RMMs) by point-to-point optical gigabit Ethernet links.
Each FPN serves up to two back-end boards.
The FPN is implemented as three tasks running as separate processes, 
interconnected by shared memory data buffers and communicating via standard inter-process mechanisms.
Readout and configuration of the electronics and all connected hardware is provided by the readout task (RXT).
The readout is parallelised across electronics boards in a multi-threaded manner
and data is buffered for access by the data processing task (DPT).
The RXT additionally receives periodic monitoring data from the TFBs 
and external temperature sensors
which it passes to the global slow control.
The DPT performs data reduction and basic data processing. 
The DPT decodes the TFB raw data blocks,
associates amplitude and timing information for individual hits,
performs pedestal subtraction on a channel-by-channel basis,
applies zero-suppression to the unsparsified data 
and formats the data for output.
To preserve monitoring information,
the DPT also performs per-channel histogramming of signal amplitudes for specific trigger types prior to zero suppression
and periodically inserts the histograms into the output data stream.
The processed event fragments are buffered and dispatched to the DAQ back-end by the third process,
which implements the MIDAS front-end functionality. 

The MIDAS framework provides the core components necessary for the DAQ system
and is used to gather the event fragments from the SMRD FPNs (and from the other ND280 sub-detectors).
An event building process collects the fragments,
performs basic consistency checks 
and writes the fully assembled events to a system buffer for output by a logger process to a local RAID array.
A custom archiver process transfers completed files to a mass storage facility over the network
and additionally makes a preview copy to a local semi-offline system for fast-turnaround analysis.

\subsection{Data Monitoring}
A custom online-monitoring (OM) server based around the ROOT framework
retrieves built events from the system buffer and generates a range 
of plots for data and detector quality monitoring as well as diagnostic alarms.
The online monitoring allows to perform real time and near online data checks 
of the SMRD. 
The ND280 online monitoring framework is derived from the consumer
monitor package developed for the Collider Detector at Fermilab (CDF) 
\cite{cdf-online} and 
is integrated into the ND280 software framework and directory structure.
The data monitoring system receives events from the data acquisition (DAQ)
system. It analyzes the data stream of events and outputs results in the
form of histograms, tables, and warnings.

The OM system consists of several data monitoring programs, one for each 
sub-detector of the ND280.
Each of the programs connects to the DAQ and continually listens to a socket. 
Upon receipt of a set number of events or an end of run message it performs
a diagnostic analysis of which the resulting histograms are saved to
a file. The monitor functions are divided into 5 different levels according to 
the number of events and frequency required to perform the checks. Level 1
monitor functions act on an event by event level whereas level 5 functions
accumulate data for one hour before being activated.

Multiple ROOT based viewer programs can connect from remote machines through
a socket connection to a display server which has access to the information 
in the shared memory in real time. Using the display program, users
can select and view multiple histograms of interests.
For the SMRD the key parameters that are being monitored are the MPPC gain, 
dark rate and position and width of individual photo electron peaks for each 
channel which are derived from multi-Gaussian fits to the respective ADC 
spectra. Data are displayed in an intuitive and accessible format for user to
be able to quickly identify problems.
An error handler process receives warning and error messages from different 
monitoring programs, displays them with appropriate action required and 
has the ability to keep detailed logs. Alert range boundaries can be adjusted
by means of a parameter settings file and without affecting the
running of the monitor.

\section{Detector Calibration and Performance}

The calibration of the SMRD response is applied in two stages.
The lower level calibration operates on individual photosensor 
signals and consist of
procedures common for all Trip-\textit{t} based subsystems. The
higher level calibration involves multi-channel,
SMRD-specific operations which are presented in Sec.~\ref{sec:zposreco}.\\
The low level calibration translates ADC signal amplitudes into
charge values in units of photo-electron equivalent (p.e.). 
Both raw ADC values, from the low- and the high-gain
output of the Trip-\textit{t} chip, are processed. After pedestal
subtraction the two ADC values are combined with a polynomial
function into a single, linearized ADC\textit{lin} value.
The ADC\textit{lin} values are scaled according to the channel gain to obtain
the charge amplitude in units of p.e. Pedestal and gain calibration
constants are updated every 3 hours for all readout channels and
stored in an online database which is common to all
Trip-\textit{t} subsystems.

\begin{figure}[htp]
\begin{center}
\includegraphics[width=0.45\textwidth]{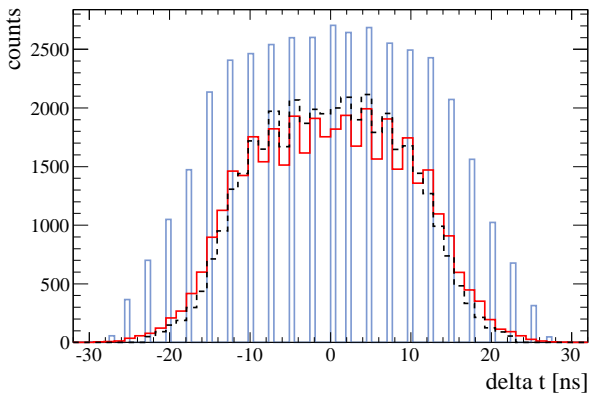}
\caption{Signal time difference distribution without timewalk
correction (solid blue, $\sigma = 11.9$ ns), with electronics
timewalk correction (solid red, $\sigma = 9.5$ ns) and with full
electronics and scintillation timewalk correction (dashed black,
$\sigma = 8.8$ ns).} 
\label{fig:sigtimingelxtw}
\end{center}
\end{figure}

Signal timing is obtained from the TDC output of Trip-\textit{t} chips.
Times correspond to the time an ADC signal surpases a 4.5 p.e. threshold.
The timing value in nanoseconds is calculated as $t = 2.5 \times TDC~bins$
and aligned with a common reference time for all subsystems.
Two charge-dependent effects influence the signal timing
(\textit{timewalk effect}). The first is related to the
Trip-\textit{t} board electronics and arises from the exponential
charging of the capacitors. It is corrected at the lower level of the 
calibration. 
The distribution of the time difference ($\Delta t$) between 
signals observed at opposite ends of a counter is shown in figure 
\ref{fig:sigtimingelxtw}.
The distribution narrows from 12 ns to 9 ns after the correction.
The second effect is caused by the light propagation in the 
scintillator and the WLS fiber which exhibit an exponential behavior.
It causes a statistical signal timing delay in
registering $n$ photons out of a total number of $N$ photons.
The expected mean value of the delay is subtracted from the signal time
and leads to a modest further reduction of the width of the distribution of 
time differences between signals at opposite ends 
(Fig.~\ref{fig:sigtimingelxtw}).

\subsection{Counter Light Yield}

\begin{figure}[htp]
\begin{center}
\includegraphics[width=0.44\textwidth]{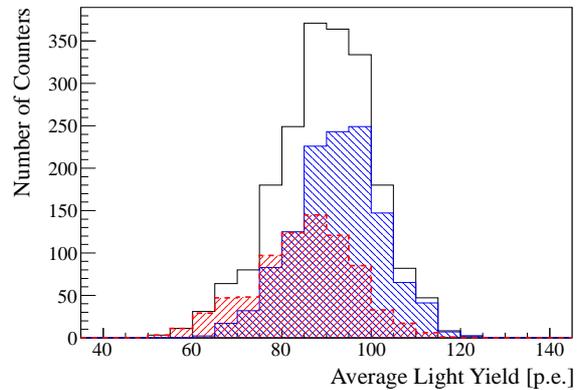}
\caption{Mean light yield of horizontal and vertical SMRD counters in 
response to highly selective set of near vertical and near horizontal cosmic 
muons. Blue and red hatched histograms represent vertical and horizontal counters, 
respectively and the black histogram corresponds to all counters.}
\label{fig:tbvert_lrhoriz}
\end{center}
\end{figure}

The light yield of all scintillation counters is a critical performance 
parameter of the SMRD and as described in section \ref{section:qa} has also 
been used for quality assurance purposes during the manufacturing process.
In test bench setups the SMRD counters showed light yields 
from 25 to 50 p.e. in response to cosmic muons penetrating the counter center
perpendicularly to its surface.
Asymmetries in signal sizes between the two ends were less than 20\%.
After installation of the SMRD counters into the UA1 magnet, cosmic ray data 
samples are used to measure the light yield for each counter.
The light yield was determined using calibrated signals from both photosensors
of each counter. The calibration corrects the signals for light attenuation
in the WLS-fiber. 
Figure \ref{fig:tbvert_lrhoriz} shows distributions of the mean light yield 
for horizontal and vertical counters in response to cosmic muons that are 
predominantly vertical and horizontal, respectively.
The cosmic muons have a near uniform distribution across the counter surface.
The blue and red hatched histograms represent vertical and horizontal 
counters, respectively. The black histogram represents all counters.
The mean value of the mean light yields amounts to 84 p.e. for horizontal 
counters and 92 p.e. for vertical counters. 
The difference in light yield for horizontal and vertical counters is 
partially due to mean muon pathlength differences in the counters caused 
by the angular distribution of cosmic muons.
The mean light yields are also larger compared
to the pre-installation test stand measurements. 
The differences can be attributed to the correction for light attentuation in the WLS fiber and differences in the counter illumination.
All counters show good performance in response 
to uniformly distributed and near perpendicular muons. 
The stability of the light
yield versus time is discussed in section \ref{stab-perf}.

\subsection{Timing Properties and Position Resolution}
\label{sec:zposreco}

For each ADC signal with an amplitude larger than 4.5 p.e. an arrival time is 
recorded.
The double sided readout of the SMRD counters therefore allows to determine 
the relative time difference between signals at each end of the counter and 
thus reconstruct the position of the particle initiating the scintillation
light in the direction of the beam.
Hit coordinates in the direction perpendicular to the beam, $x_{hit}$ and
$y_{hit}$ are always set to the middle of the scintillator slab since 
there is no position sensitivity in these directions. 

\begin{figure}[htp]
\begin{center}
\includegraphics[width=0.45\textwidth]{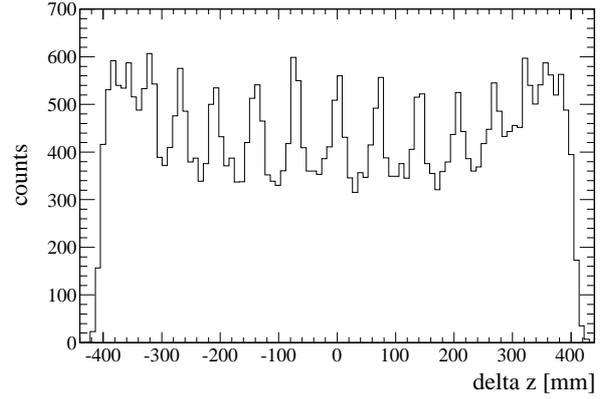}
\caption{Distribution of estimated hit position $z_{hit}$
for horizontal counters and cosmic muon data.} \label{fig:hitzpos}
\end{center}
\end{figure}

Determination of the particle hit time ($t_{hit}$) and successive position
reconstruction along the counter ($z_{hit}$, e.g. in the direction of the 
beam) requires a signal at each end of the counter.
Both signals are required to be above a threshold amplitude of 
4.5 p.e. and within a 23 ns time coincidence window. 
The size of the coincidence window was chosen 
to minimize contamination from 
 accidental noise coincidences without rejecting a significant
fraction of good hit candidates.
The estimated light propagation time for a signal to travel end to end in the
embedded WLS fiber is 13 ns. Hence the chosen coincidence window
leaves a margin for uncertainty in the timing measurement of individual 
signals (see Fig.~\ref{fig:sigtimingelxtw}).
The particle hit time is obtained from the times of individual signals as
\begin{equation}
t_{hit} = t_{qmax} - \frac{l_z}{2v_{eff}} - \frac{1}{2}\left(\Delta t
\right),
\end{equation}
where $t_{qmax}$ is the arrival time of the larger of the two signals,
$l_z$ is the length of the counter slab (876 mm),
$v_{eff}$ is the effective speed of light along the counter axis which is
measured to be 65.5 mm/ns and $\Delta t$ is the time difference between 
the arrival times of signals at both ends.\\
The z-hit position is evaluated w.r.t. the middle of the counter, 
and is calculated based on the signal timing difference 
and the charge asymmetry of the two signals.
A two-stage Bayesian estimate is applied to combine the time and charge 
information optimally. 
A uniform prior probability distribution of the particle hit position 
is assumed since cosmic muons are expected to illuminate the counters uniformly.
Successively the probability distribution for the particle hit position is
constructed in two consecutive steps which are based on 
charge asymmetry and timing difference measurement information. 
The uncertainties of the time and charge measurements are
taken into account as likelihood parameters.
An evaluation of the probability distributions for arrival times and signal 
charges
is implemented numerically in the off-line calibration software and results in 
a high granularity in hit position.
The particle hit position bias and position uncertainty are taken from
the resulting probability distribution as the mean value and
standard deviation. 
Results of the position
reconstruction for horizontal counters and cosmic muon data are shown
in Fig.~\ref{fig:hitzpos}. The peak structure is
caused by signal timing quantization.
 The resolution of the particle hit position
reconstruction is at the level of 85 mm.

\subsection{Counter efficiencies}
The efficiencies of individual SMRD counters has been studied using cosmic 
trigger data which are recorded in between beam spills and in dedicated 
cosmic runs while the beam is off. For the purpose of measuring particle detection efficiencies SMRD counters are divided into two groups, those which 
are sandwiched by other counters and counters which are in the innermost or outermost radial layers.
Efficiencies of counters in the first group can be measured by tagging 
penetrating cosmic muon with the bracketing counters on either side and 
determining the fraction for which a muon was observed in the bracketed layer.
The tagging requires each of the two bracketing counters to observe a double 
ended signal, with individual signal amplitudes larger than 4.5 p.e. and 
within a coincidence window of 23 ns.
The signals of both bracketing counters 
have to be within a time window of 70~ns to provide a trial event. 
The ratio of observed events in the sandwiched counter to the total number of 
trial events is recorded as the detection efficiency of that particular counter.
The mean particle efficiency for the 856 counters in category one was found to 
be 97.0 $\pm$ 0.4\% for horizontal and 97.9 $\pm$ 0.2\% for vertical counters. 
For counters in the second category a somewhat 
more sophisticated analysis including track reconstruction is required in order
to estimate whether a muon penetrated a given counter in the inner or outermost
radial layer. Studies find the detection efficiency for counters in the second category to be comparable to those in the first category.
Counter efficiencies are monitored as function of time and are found to be 
stable within uncertainties.

\subsection{Detector Stability and Performance}
\label{stab-perf}

\begin{figure}[htp]
\begin{center}
\includegraphics[width=0.45\textwidth]{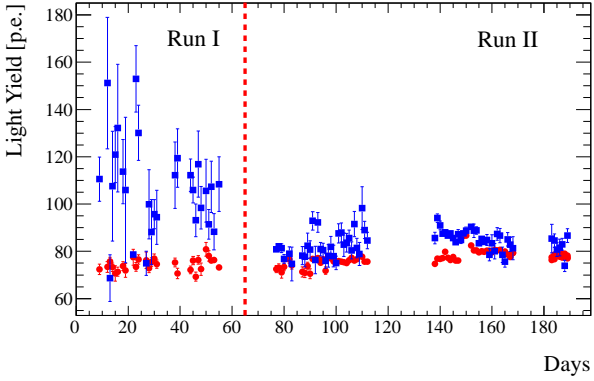}
\caption{Mean light yield versus time for horizontal (red circles) and vertical (blue squares) SMRD counters in response to nearly perpendicularly penetrating cosmic muons.}
\label{fig:cosmic_stability}
\end{center}
\end{figure}

The performance of the SMRD is being monitored by recording the light yield
in response to the constant flux of cosmic muons which penetrates the detector. 
Cosmic trigger data are collected in between beam spills 
throughout the entire data taking period and during dedicated 
cosmic muon runs at times when the beam is off.
Furthermore, beam data can be used to monitor the SMRD performance if 
the bunched beam signals are normalized by the number of protons on target (POT)
and the uncertainties on the beam are measured by other detectors. 
Alternatively, if
the stability and response of the SMRD is well characterized with cosmic trigger data the SMRD data can be used to provide valuable cross checks on the neutrino beam stability. 
The present section demonstrates the SMRD performance based on cosmic muon 
and pre-selected beam data.
For the present analysis two cosmic muon data samples were selected,
one sample of muons which are vertical to within 10$^\circ$ 
and a second sample of muons which are horizontal to within 10$^\circ$.
Figure \ref{fig:cosmic_stability} shows good stability in the
mean light yield for the vertical and the horizontal cosmic muon samples as function of time. The larger and less stable light yield for vertical counters  
during run I is caused by varying trigger settings for the collected data 
sample of horizontal cosmic muons.

\begin{figure}[htbp]  
\begin{center}                                                                  
\includegraphics[width=0.45\textwidth]{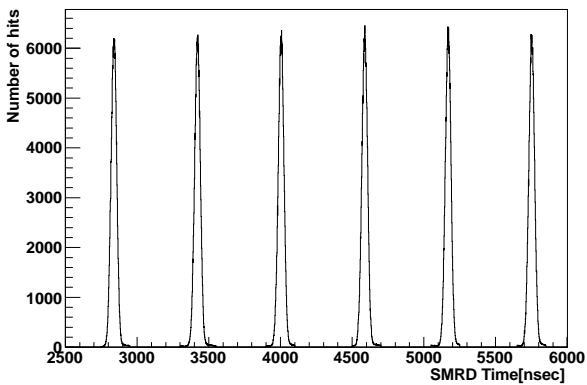}         
\caption{Beam triggered hit time distribution as recorded by the SMRD during run 1 for 2.94$\times 10^{19}$ POT.}

\label{fig:beamtiming}                                                 
\end{center}                                                                    
\end{figure}

\begin{figure}[htp]
\begin{center}
\includegraphics[width=0.45\textwidth]{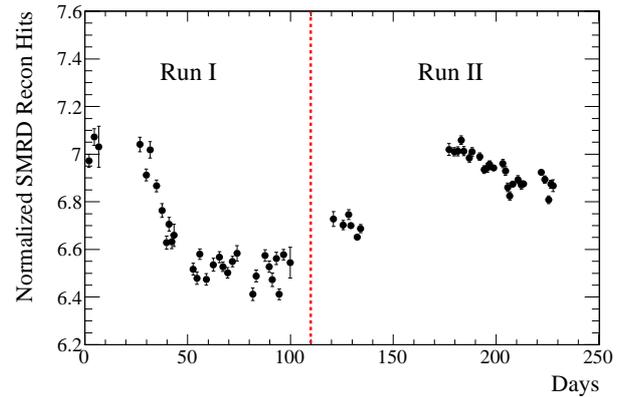}
\caption{SMRD counter Recon hits 
normalized by POT and versus time in units of days.
The units of the vertical axis are Recon hits per 1$\times$10$^{14}$ protons 
on target. The error bars reflect statistical uncertainty only.}
\label{fig:evt-stability}
\end{center}
\end{figure}

During the first two run periods from January 2010 to March 2011
the J-PARC accelerator and neutrino beamline delivered
1.45$\times 10^{20}$ protons on target. 
After applying beam data quality criteria, a data
set corresponding to 1.43$\times 10^{20}$ POT was identified.
The combined duty factor of the SMRD and its data acquisition amounts to 0.97
for the above data taking period.
The 6 bunch beam structure used in run 1 (8 bunches for run 2) is clearly visible in the data and displayed in figure \ref{fig:beamtiming} which is for
an integrated total of 2.94$\times 10^{19}$ POT.
Shown are SMRD counter hits with a double-sided coincidence. Individual hits 
contributing to the coincidence are required to have signal amplitudes larger 
than 10 p.e. and fall within a coincidence window of 23 ns.
The electronics integration windows are 480~ns wide and their timing has been 
adjusted to accept bunch signals in the earlier part of the window. Hence, 
the later part of each integration window shows random backgrounds as well as 
a decreasing tail which can be attributed to Michel electrons originating 
from muon decay.
Each integration window is followed by a 100~ns dead time of the electronics and 
the nominal bunch spacing is 581~ns.
Beam data collected from March 2010 until March 2011 are used to monitor the
SMRD stability over time. 
The analysis uses double-sided SMRD counter hits (Recon hits) which 
are composed of
single hits on either side of the counter with a charge amplitude greater 
than 4.5 p.e. and within a 23~ns coincidence window. In addition, the hits are
required to fall within a 200ns time window centered on the beam bunches.
The rate of background events is estimated from the electronic integration
cycles which do not coincide with a beam bunch and is found to be negligible. 
Nonetheless, background events are subtracted from Recon hits 
before these are normalized to
protons on target (POT). The normalized Recon hits are corrected for 
temperature and plotted versus time.
The temperature dependence of the Recon hit rate is due to the sensitivitity
of the MPPC overvoltage to temperature as described in \cite{mppc}.
Fig. \ref{fig:evt-stability} shows the background subtracted SMRD Recon hits
normalized to POT and corrected for temperature 
effects as function of time.
The normalized SMRD Recon hit rate is found to be stable to within 10\%.

\begin{figure}[htb]
\begin{center}
\includegraphics[width=0.45\textwidth]{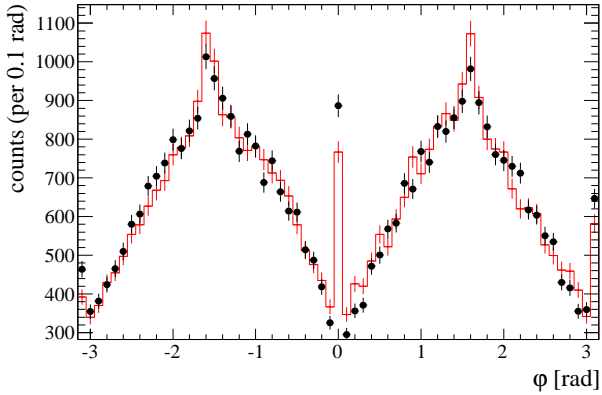}
\caption{Comparison of the distributions of the azimuthal angle $\phi$
in data (black points) and simulation (red histogram) for an unbiased cosmic trigger.}
\label{fig:cosmic-comp-phi}
\end{center}
\end{figure}

\begin{figure}[htb]
\begin{center}
\includegraphics[width=0.45\textwidth]{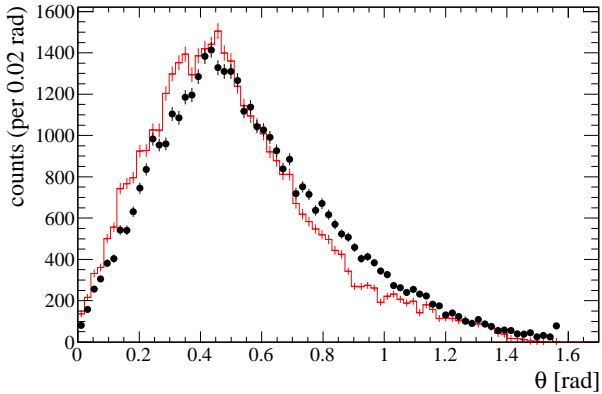}
\caption{Comparison of the distributions of the zenith angle $\theta$
based on data (black data points) and simulation (red histogram) for
an unbiased trigger configuration.}
\label{fig:cosmic-comp-theta}
\end{center}
\end{figure}

The SMRD response to cosmic muons has been cross checked with
a simulated data set of cosmic muons.
The simulation assumes a realistic zenith angle and momentum distribution 
of cosmic muons on the surface of Earth.
It includes the propagation of muons for zenith angles up to 85$^\circ$ 
through sand in order to reach the detector, the passage of muons through 
the detector and the creation of electronic hits. The last stage of the
simulation is the application of the trigger conditions described in
section \ref{chapter:triggerlogic}. In the simulation the efficiency of the 
detector and the trigger system was assumed to be 100\%.
The simulated cosmic muon sample is submitted to the same calibration and 
reconstruction procedure as the data. 
A comparison of the angular distributions for data and the simulated
sample is shown in Fig.~\ref{fig:cosmic-comp-phi} and
\ref{fig:cosmic-comp-theta} 
for an unbiased cosmic ray sample in which
all SMRD towers contribute equally to the formation of a trigger.
The azimuthal distribution for reconstructed cosmic muon data and 
corresponding simulation agree within uncertainties.
For the zenith angle distribution the reasonable agreement between data and 
simulation is found and can serve as an indication of the level of 
understanding of the detector response.

\section{SMRD Event Reconstruction}

A particle track reconstruction in the SMRD is incorporated 
into the ND280 software package which is designed for physics analyses of 
ND280 data and to perform detector simulations.
The SMRD specific software components consist of a SMRD response
simulation, a calibration package and a track reconstruction package 
solely based on SMRD hit information. Furthermore, a global reconstruction
package allows to reconstruct particle interactions inside the detector based
on information from all ND280 sub-detectors.
The present section gives a brief overview of the event reconstruction algorithms
which are based on SMRD information.

The SMRD reconstruction is based on \textit{Recon hits} ("reconstructed hits") in the SMRD which are built based on time and charge signal information from both sides of the counter and which take the coordinate in the long direction of the counters into account. 
	The main algorithm used to incorporate SMRD information into the global ND280 reconstruction is a filtering of hits on a hit-by-hit basis.  An adjacent track from the ND280 inner sub-detectors is used as a seed and hits from each SMRD wall are tested against it using a Kalman filter \citep{kalman}. Kalman filtering is performed with the RecPack reconstruction toolkit \citep{recpack} which handles navigation in magnetic fields, energy loss corrections, multiple scattering and similar processes.  The algorithm enables to match even single SMRD hits with high efficiency. A complementary algorithm is the RecPack based matching of inner detector tracks with SMRD tracks which were reconstructed within the
SMRD only.
	
The principal purpose of event reconstruction based solely on SMRD information
is to reconstruct tracks from neutrino interactions in the SMRD and which are not matched to any inner detector activity. For beam induced events SMRD activity in opposing sides of the SMRD does generally not originate from single muons.
An additional goal is to identify and reconstruct cosmic muons which are used for calibration and detector monitoring purposes. 

Since the maximum number of scintillator layers in a tower of the SMRD is six
the SMRD-only reconstruction of tracks in one side of the SMRD is based on a small number of hits. In general, a straight line fit is a good approximation
to the true particle track in the SMRD. The reconstruction starts with the identification of clusters of hits which is done using hit time stamps and geometrical neighbor criteria. Track fitting is based on the Principal Component Analysis (PCA) \citep{pca} method, of which the first component is taken as a fit. For further usage in the global ND280 event reconstruction, SMRD tracks are refitted with the RecPack based Kalman filter in order to obtain covariances. 

Individual cosmic muons can produce hits on multiple sides of the SMRD and 
are used
for calibration and performance monitoring purposes. The associated
long muon tracks across the ND280 detector are 
fitted by a PCA method that incrementally removes hits which are located 
sufficiently far from the initially fitted track. 
In addition, the possibility exists to match track segments in different SMRD 
quadrants to form a single long track.

\begin{figure}[htb]
\centering\includegraphics[width=0.5\textwidth]{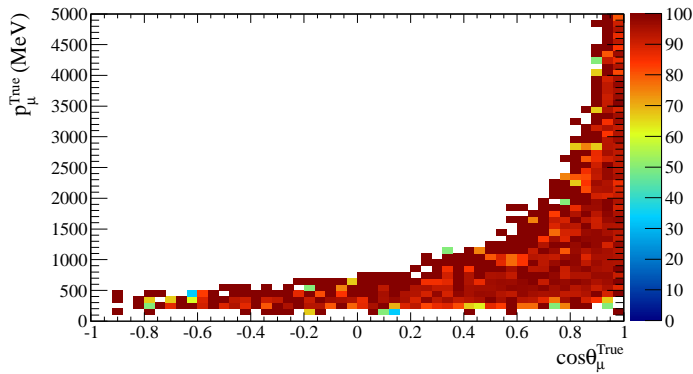}
\caption{Track finding efficiency of SMRD-only reconstruction for events with at least two muon hits in the detector as a function of true muon momentum and cosine of the angle with respect to the beam direction.}
\label{fig:smrdReconEff}
\end{figure}

The track finding efficiency of the SMRD only reconstruction algorithm as a function of true muon momentum and cosine of the angle with respect to neutrino beam direction is shown in Figure \ref{fig:smrdReconEff}. Neutrino interactions have been simulated within the entire ND280 detector volume. Neutrino energies are derived from a full simulation of the neutrino beam flux \cite{t2knim} and events follow the time structure of the beam. Neutrino interactions are handled by the software package NEUT \cite{neut} and secondary particles are propagated and converted to time and charge signals by the custom built GEANT 4 based ND280 software.

\section{Summary and Conclusion}
The elements of the SMRD of the T2K near detector have been described.  
The construction of the SMRD was completed in summer of 2009 and
followed by a successful commissioning phase. 
The performance of the SMRD was evaluated with cosmic ray data and
found to meet or exceed specifications. The detector response to
cosmic muons was analyzed and found to agree with simulations.
T2K started taking neutrino beam data for physics analysis in January 2010.

\subsection{Acknowledgements}

The authors thank 
the Department of Energy, U.S.A.;
the Russian Academy of Sciences, the Russian Foundation for Basic
Research, and the Ministry of Education and Science of the
Russian Federation, Russia;
the Polish National Science Centre, Poland; and
the Science and Technology Facilities Council, U.K. 
In addition, the participation of individual researchers and LSU in
the construction of the SMRD has been further supported by funds from 
the U.S. Department of Energy Outstanding Junior Investigator (OJI) Program.






\bibliographystyle{model1a-num-names}
\bibliography{<your-bib-database>}







\end{document}